\documentstyle[prb,aps,eqsecnum,multicol,epsf]{revtex}
\renewcommand{\narrowtext}{\begin{multicols}{2} \global\columnwidth20.5pc}
\renewcommand{\widetext}{\end{multicols} \global\columnwidth42.5pc}
\newcommand{\Lrule}{\vspace*{-0.2in}\noindent\vrule width3.5in height.2pt
  depth.2pt \vrule depth0em height1em}
\newcommand{\Rrule}{\vspace{-0.1in}\hfill\vrule depth1em height0pt \vrule
  width3.5in height.2pt depth.2pt\vspace*{-0.1in}}
\begin{document}
\draft

\title{Random matrices and the replica method}
\author{E. Kanzieper}
\address{The Abdus Salam International Centre for Theoretical Physics, 
P. O. Box 586, 34100 Trieste, Italy}
\date{29 August 1999}
\maketitle

\begin{abstract}
Recent developments [Kamenev and M\'ezard, cond-mat/9901110, cond-mat/9903001;
Yurkevich and Lerner, cond-mat/9903025; Zirnbauer, cond-mat/9903338] have revived a 
discussion about applicability of the replica approach to description of spectral
fluctuations in the context of random matrix theory and beyond. The 
present paper, concentrating on invariant non--Gaussian random matrix ensembles
with orthogonal, unitary and symplectic symmetries, aims to demonstrate that both
the bosonic and the fermionic replicas are capable of reproducing nonperturbative
fluctuation formulas for spectral correlation functions in entire energy scale,
including the self-correlation of energy levels, provided no $\sigma$--model 
mapping is used. 
\end{abstract}
\narrowtext
\section{Introduction}\setcounter{equation}{0}

Energy level statistics of electrons in a random potential is one of the basic 
issues in the theory of disordered conductors. It is also a classic example of
the problem whose proper statistical description calls for a use of nonperturbative
methods of averaging over disorder realizations. The latter procedure
is not trivial because the spectral observables depend in highly nonlinear fashion on a
random Hamiltonian, thus making the calculation of the ensemble averages very difficult.
Within the single electron picture, it is necessary to average a product of resolvents,
$\left( {\bf r}|( z - {\cal H})^{-1}|{\bf r'}\right)$, that admits an
exact representation through a ratio of two functional integrals over auxiliary
fields which may consist of either commuting [bosonic] or anticommuting [fermionic]
entries. This representation, however, is still not convenient for nonperturbative
averaging due to an awkward denominator. To get rid of it, Edwards and Anderson
\cite{EA-1975} proposed a replica method that substitutes the initial disordered system 
by the $n$ identical noninteracting systems. Such a trick removes a randomness from the denominator
and, therefore, makes it possible to perform a nonperturbative averaging over the random
potential from the very beginning for all integer $n$ which must be set to zero at the
appropriate stage [by the analytic continuation $n\rightarrow 0$]. Depending on the
origin of the auxiliary fields, the resulting effective field theory is defined on
either a noncompact \cite{W-1979} or a compact \cite{ELKh-1980} manifold.

An alternative approach was invented by Efetov \cite{E-1982} who introduced 
an auxiliary field with an equal number of bosonic and fermionic entries that obviates
the need for replicas. This supersymmetry formalism turned out to be very fruitful in
the theory of disordered conductors and beyond \cite{E-1997}. Along with other fundamental
results, it has led to establishing a link \cite{E-1982,VWZ-1985} between the theory of 
disordered metals and the random matrix theory [RMT] founded by Wigner 
\cite{W-1951}, thus confirming on microscopic
grounds a much earlier conjecture by Gorkov and Eliashberg \cite{GE-1965}. To be precise, 
it was demonstrated \cite{E-1982} that the `density--density' correlation function $R_{\beta}(s)$ of disordered conductor in the ergodic limit is 
solely determined by the global symmetries of the system and coincides with the
RMT predictions. For the three symmetry classes -- orthogonal $(\beta =1)$, unitary
$(\beta = 2)$, and symplectic $(\beta =4)$ -- the `density--density' correlation functions
are
\widetext
\Lrule
\begin{mathletters}
\label{tlcf-susy}
\begin{eqnarray}
\label{eq01}
R_{1}(s) &=& 1 - \frac{\sin {}^2\left( \pi s \right) }
{\left( \pi s \right) ^2} - 
\int_{s}^{+\infty} dt \frac{\sin(\pi t)}{\pi t} \frac{d}{ds}
\left( \frac{\sin (\pi s)}{\pi s}\right) + \delta(s),
\label{orth} \\
R_{2}(s) &=& 1-\frac{\sin {}^2\left( \pi s \right) }
{\left( \pi s \right) ^2}+\delta \left( s \right),
\label{unit} \\
R_{4}(s) &=& 1-\frac{\sin {}^2\left( 2\pi s \right) }
{\left( 2\pi s \right) ^2}+ \int_{0}^s dt \frac{\sin(2\pi t)}{2\pi t} 
\frac{d}{ds}
\left( \frac{\sin (2\pi s)}{2\pi s}\right) + \delta(s),
\label{sympl} 
\end{eqnarray}
\end{mathletters}
\Rrule
\narrowtext
\noindent
$s$ being the distance between the eigenlevels measured in units of the
mean level spacing $\Delta$. The range of validity of 
Eqs. (\ref{tlcf-susy})
is $s \ll g$, where $g\gg 1$ is the dimensionless Thouless conductance.

While the success \cite{E-1997} of the supersymmetry approach has been very impressive, 
the alternative replicated field theories \cite{W-1979,ELKh-1980} [which, besides, are 
suitable for a treatment of disordered systems in presence of interactions 
\cite{F-1983}] seemed to be restricted \cite{VZ-1984} to reproducing the perturbative diagrammatic
expansions \cite{AS-1986} only. A detailed analysis of this problem has been given by
Verbaarschot and Zirnbauer in Ref. \cite{VZ-1985}, where the failure of the replica method
[in its $\sigma$--model formulation] to 
correctly account for all nonperturbative contributions has been attributed
to the fact that the replica trick is mathematically ill founded, and the 
knowledge of the replicated partition function for all integer $n$ need not be sufficient
for extrapolation to the limit $n\rightarrow 0$ which, generically, can also be 
nonunique \cite{HP-1979}.

Surprisingly enough, in the very recent publication \cite{KM-1999a}, Kamenev and M\'ezard 
have formulated a variant of the fermionic replica 
method that allowed them to rederive Eq. (\ref{unit}), $s \gg 1$, in the context of RMT for the Gaussian Unitary Ensemble
[GUE]. The approach \cite{KM-1999a} has relied on the combination of the exact integration 
over 
the angular degrees of freedom of the replicated $\sigma$--model, by using the 
Itzykson--Zuber integral \cite{IZ-1980}, and the approximate saddle point calculations 
in the $n$--dimensional space of remained eigenvalues 
$\hat{\Lambda} = {\rm diag}(\lambda_1,\ldots,\lambda_n)$. In order to correctly reproduce
the result Eq. (\ref{unit}) for GUE in the large--$s$ limit, one
is forced to take into account the nontrivial saddle points which break the 
replica symmetry between the different eigenvalues $\lambda_i$, $1 \le i \le n$. On the formal
level, this 
implies that 
the $n$--component vector of eigenvalues ascribes a nontrivial internal structure which 
must be kept in the limit $n\rightarrow 0$ of vanishing number of integration variables! 
Very similar structure of the vector saddle point manifolds is present in two subsequent 
publications \cite{KM-1999b,YL-1999} that extended the results of Ref. \cite{KM-1999a} to 
two other [orthogonal and symplectic] symmetry classes both in the RMT limit \cite{YL-1999}
and beyond it \cite{KM-1999b,AA-1995}. A critical discussion of the procedures employed
in Refs. \cite{KM-1999a,KM-1999b,YL-1999} has been given in Ref. \cite{Z-1999} where
it has been shown that there exists some additional input in the calculational schemes
\cite{KM-1999a,KM-1999b,YL-1999} which is equivalent to introducing the causality
information. A conjecture was put 
forward \cite{Z-1999} that the fermionic replicas in elaboration 
\cite{KM-1999a,KM-1999b,YL-1999} are 
perturbatively equivalent to the supersymmetry method, with $s^{-1}$ being a small 
parameter. It was also explained \cite{Z-1999} that the results \cite{KM-1999a,YL-1999} 
are `semiclassically exact' for the unitary symmetry class owing to the 
Duistermaat--Heckman theorem \cite{DH-1982}.

There is one common feature inherent to the above mentioned replicated
calculations: All of them, 
starting with either
bosonic or fermionic replicas, reduce the problem of evaluating the spectral 
density--density correlation function to the effective $\sigma$--model defined in the
configurational space of replicas only. Correspondingly, the `coordinate' space 
disappears from the theory, and the memory of the initial [matrix] Hamiltonian is only
retained in the symmetry of the auxiliary $\hat {Q}$--fields. An immediate consequence of 
such a treatment is that the integral representations 
for the spectral observables appear to involve the integration measures which 
become senseless as $n\rightarrow 0$. This can be seen, for instance, from the 
Verbaarschot--Zirnbauer integral for the two--level correlation function derived 
\cite{VZ-1985} from the fermionic replicas for the GUE in the large--$N$ limit:
\begin{eqnarray}
\label{vzi}
R(r) &=& - \lim_{n \rightarrow 0} \frac{1}{n^2} \partial_r^2 \int_{-1}^{+1}
\prod_{m=1}^{n} \left( d\lambda_m {\rm e}^{i r \lambda_m}\right) \nonumber \\
&\times& \prod_{1 \le m_1 < m_2 \le n}
\left|\lambda_{m_1} - \lambda_{m_2}\right|^2.
\end{eqnarray}
[Here, $r=\pi s$ in notations of Eq. (\ref{unit})]. Obviously, one cannot simply
take the limit $n\rightarrow 0$ neither in the above integral representation nor in 
the result of the integration
\begin{equation}
\label{vzi-det}
R(r) = - \lim_{n\rightarrow 0} \frac{1}{n^2} \partial_r^2 \det
\left( \partial_r^{i+j-2} \frac{\sin r}{r}\right)_{1 \le i,j \le n}
\end{equation}
because of the vanishing dimensionality of the matrix under the sign of the determinant. 
Therefore, one should know how to evaluate the integrals of the type 
Eq. (\ref{vzi}) in the closed form for arbitrary integer $n$ in order to analytically
continue the resulting expression from integer $n$ and perform the replica limit
$n \rightarrow 0$ afterward. One such way proposed 
in Ref. \cite{YL-1999}
reduces Eq. (\ref{vzi}), in the domain $r\gg 1$, to the expression
which coincides identically with the one obtained by means of the `replica symmetry
breaking' [RSB]. This coincidence signals that the RSB may be viewed as a 
useful language for approximate computation of the integrals defined on the problematic [in the above
mentioned sense] measures. In the course of exact calculations, the notion of RSB must not appear
at all. 

The studies \cite{VZ-1985,KM-1999a,KM-1999b,YL-1999,Z-1999} call for further clarification
of the question that has a conceptual importance and concerns the applicability of the 
replica method to a nonperturbative
description of the eigenlevel correlations in general, and those on the {\it short 
distance}
scale, in particular. The central issue to be learnt here is whether the replica method
{\it itself} faces some internal insurmountable obstacles, or existing difficulties 
are to be attributed to the particular computational schemes. We believe that in order 
to answer this question, one has to depart from conventional routes toward exact 
calculational schemes within the replica framework. 

While it is unknown so far how the exact replica treatment could be realized for 
realistic disordered systems, the present paper aims to demonstrate that
there is one particular but conceptually important situation where the exact computation 
is possible for both the fermionic and the bosonic replicas. This
is the case of invariant random matrix ensembles \cite{M-1991} with a general non--Gaussian measure
\begin{equation}
\label{ngue}
P_N[{\bf H}] \propto e^{-\beta{\rm tr} V[{\bf H}] }
\end{equation}
specified by the confinement potential $V[{\bf H}]$ and the Dyson index $\beta=1,2$ and
$4$ that labels the symmetry of the matrix Hamiltonian ${\bf H}$.

A distinctive feature of our treatment that must be emphasized from the 
very beginning is that we do {\it not} use a $\sigma$--model representation 
for the replicated partition function averaged over the ensemble Eq. (\ref{ngue})
of random matrices. Instead, a different technique is used for exact averaging of the 
product statistics that comprises the replicated partition function as a 
particular case. This technique generates especially useful representations for the
spectral observables, thus allowing one to avoid the problem of the ill defined 
integration measure. As a direct consequence of that, we construct [in a simple and 
straightforward way] a well defined replica limit 
$n \rightarrow 0$ in expressions for the spectral observables that turn out to 
identically coincide with Eqs. (\ref{tlcf-susy}). This coincidence, observed
for arbitrary $s$, is the main outcome of the study.

The following program is carried out:

(i) First, the replicated partition function of the form
\begin{eqnarray}
Z^{n}_{\gamma, \gamma^{\prime}}(E,E^{\prime}) &=& 
{\rm det}^n\left[ (E-{\bf H})^2 + \gamma^2\right] \nonumber \\
&\times& {\rm det}^n\left[ (E'-{\bf H})^2 + \gamma^{\prime 2}\right]
\label{rpf-intro}
\end{eqnarray}
is introduced in accordance with the very concept of the replica method that
boils down to constructing such a generating function [e. g., for the `density--density'
correlator] which does not contain a random denominator at the expense
of the artificial appearance of replica parameter $n$ in Eq. (\ref{rpf-intro}).

(ii) Second, the replicated partition function is averaged over the ensemble 
Eq. (\ref{ngue}) of random matrices. At this stage, we depart from Refs. 
\cite{VZ-1985,KM-1999a,KM-1999b,YL-1999,Z-1999} and
do exact averaging {\it without} appealing to a nonlinear $\sigma$--model. This is
the most crucial step in our derivation: It allows us to retain control over the 
analyticity of the resulting expressions considered as a function of $n$, the
number of replicas.

(iii) Third, the `density--density' correlation function is recovered by means of the
double limiting procedure \cite{Remark-unitary}
\begin{eqnarray}
\langle \rho_N(E) \rho_N(E^{\prime}) \rangle &=& \lim_{\gamma, \gamma^{\prime} \rightarrow 0}
\lim_{n \rightarrow \pm 0} \frac{1}{(2\pi n)^2} \nonumber \\
&\times& \partial_\gamma \partial_{\gamma ^{\prime}}
\langle Z^{n}_{\gamma, \gamma^{\prime}}(E,E^{\prime}) \rangle 
\label{limit-dos-dos}
\end{eqnarray}
and found to follow Eqs. (\ref{tlcf-susy}) in the spectrum bulk after appropriate
rescaling of spectral variables. [The limits $n\rightarrow \pm 0$ correspond to the 
fermionic and the bosonic replicas, respectively.]

The paper is organized as follows. In the auxiliary Section II, the notion of the product 
statistics is introduced, and a technique to average the product statistics is presented.
With minor modifications, this section closely follows the recent paper by Tracy
and Widom \cite{TW-1998} whose impact we acknowledge. The results of Section II,
represented in the form of the Grassmann functional integrals, are used
in Section III where the fluctuation formulas Eqs. (\ref{tlcf-susy}) are obtained.
Conclusions are presented in Section IV.

\section{Product statistics in random matrix theory}

Let us introduce the notion of the {\it product statistics} $F$ defined on the 
eigenvalues $\{\lambda\}$ of $N\times N$ random matrix ${\bf H}$ of appropriate
symmetry:
\begin{eqnarray}
\label{m-stat}
F(\lambda_1, \ldots, \lambda_N) = \prod_{k=1}^{N} f(\lambda_k).
\end{eqnarray}
Here, $f(\lambda)$ is an arbitrary function of the variable $\lambda$ such that the
integral $\int d\lambda \lambda^j f(\lambda) e^{-\beta V(\lambda)}$ exists for
positive integers $j$.

This section aims to establish exact formulas for the average of the product statistics
\begin{eqnarray}
\label{ms-av}
\langle F \rangle = \langle \prod_{k=1}^{N} f(\lambda_k)\rangle
\end{eqnarray}
over the distribution function Eq. (\ref{ngue}) for all three symmetry classes 
$\beta =1$, $2$ and $4$. We notice that along with utility of these formulas for
computing the average of the replicated partition function, they may find applications
in evaluating the average of other product statistics encountered in description
of a number of observables \cite{shot-noise,AS-1995} in mesoscopic physics.

\subsection{Unitary symmetry: $\beta=2$}
Due to invariance of the distribution function $P_N[{\bf H}]$ at $\beta=2$ under the unitary
transformation ${\rm U}(N)$, the average $\langle F \rangle$ can be 
written down as $N$--fold integral
\begin{eqnarray}
\label{un-2}
\langle F \rangle \propto 
\int {\cal D}{\hat {\Lambda}} e^{-2{\rm tr} V(\hat {\Lambda})} 
|\Delta_N(\hat {\Lambda})|^2 \det [f(\hat {\Lambda})],
\end{eqnarray}
where the proper normalization is to be restored on the later stage.
Here, the diagonal matrix $\hat {\Lambda} = {\rm diag} (\lambda_1,\ldots,\lambda_N)$, 
the integration measure ${\cal D}\hat{\Lambda} = \prod_{m=1}^{N}d\lambda_m$, and
$\Delta_N(\hat{\Lambda}) = \prod_{1 \le i < j \le N}(\lambda_i - \lambda_j)$
is the Vandermonde determinant.

Integration in Eq. (\ref{un-2}) can be performed if we
notice that the Vandermonde determinant $\Delta_N(\hat{\Lambda}) = \det[\lambda_i^{j-1}]
\propto \det[P_{j-1}(\lambda_i)]$, where $P_j(\lambda)$ is a set of 
polynomials that we choose to be orthonormal with respect
to the measure $e^{-2V(\lambda)}d\lambda$. In terms of the `wave functions' 
$\varphi_j(\lambda) = e^{-V(\lambda)}P_j(\lambda)$ this orthonormality is expressed as
\begin{eqnarray}
\label{wf-2}
\int d\lambda \varphi_i(\lambda) \varphi_j(\lambda) = \delta_{ij}.
\end{eqnarray}
The integral Eq. (\ref{un-2}) can be 
brought to the form
\widetext
\Lrule
\begin{equation}
\label{numer}
\int \left( \prod_{m=1}^N d\lambda_m f(\lambda_m) \right)
\det[\varphi_{j-1}(\lambda_i)]_{1 \le i,j \le N} 
\det[\varphi_{j-1}(\lambda_i)]_{1 \le i,j \le N}.
\end{equation}
Integrating the product of two determinants in 
accordance with 
the formula \cite{A-1883}
\begin{equation}
\label{formula-1}
\int \left(\prod_{m=1}^N d\lambda_m \right) \det[u_j(\lambda_i)]_{1\le i,j \le N} 
\det[w_j(\lambda_i)]_{1 \le i,j \le N}
= N! \det \left(
\int d\lambda u_i(\lambda) w_j(\lambda)
\right)_{1 \le i,j \le N}
\end{equation}
\Rrule
\narrowtext
\noindent
that is proven via the Laplace expansion of the two determinants, we
reduce Eq. (\ref{numer}) to
\begin{equation}
\label{qq}
\langle F \rangle \equiv
\det\left(
\int d\lambda f(\lambda)
\varphi_{i-1}(\lambda)\varphi_{j-1}(\lambda)
\right)_{1 \le i,j \le N}
\end{equation}
that enjoys a correct normalization owing to Eq. (\ref{wf-2}) and 
constitutes the first preliminary result to be used in the following
Section.

\subsection{Symplectic symmetry: $\beta=4$}

In the case of the symplectic symmetry, ${\rm Sp}(N)$, the average $\langle F \rangle$ is 
given by 
\begin{eqnarray}
\label{os-2}
\langle F\rangle \propto 
{\int {\cal D}\hat {\Lambda} e^{-4{\rm tr} V(\hat {\Lambda})} 
|\Delta_N(\hat {\Lambda})|^{4}
\det [f(\hat{\Lambda})]},
\end{eqnarray}
where the correct normalization is to be found yet.  
Despite of appearance of $|\Delta_N(\hat{\Lambda})|^4$, that reflects the self-dual
quaternion nature of the matrix ${\bf H}$ for $\beta=4$ in Eq. (\ref{ngue}), the exact 
integration in
Eq. (\ref{os-2}) is still possible. First, we make use of the identity \cite{S-1893}
\begin{eqnarray}
|\Delta_N(\hat{\Lambda})|^4 &=& \det\left[
\lambda_i^{j-1},j\lambda_i^{j-1}\right] \label{os-1} \\
&=& \det \left[
P_{j-1}(\lambda_i),P_{j-1}^{\prime}(\lambda_i)\right]_{1 \le i \le N, 1 \le j \le 2N}
\nonumber
\end{eqnarray}
to rewrite Eq. (\ref{os-2}) as
\begin{eqnarray}
\label{os-3}
&\int& \left(
\prod_{m=1}^N d\lambda_m {\rm e}^{-4V(\lambda_m)}
f(\lambda_m)\right) \nonumber \\
&\times&
\det \left[ P_{j-1}(\lambda_i),P_{j-1}^{\prime}(\lambda_i) \right]
_{1 \le i \le N, 1 \le j \le 2N}.
\end{eqnarray}
In both equations above, $P_j(\lambda)$ is a
set of polynomials which we choose to be orthonormal with respect to the
measure $e^{-4V(\lambda)}d\lambda$.  Second, the integration in Eq. (\ref{os-3}) can be
performed via the analog \cite{dB-1955} of Eq. (\ref{formula-1}) that reads
\widetext
\Lrule
\begin{equation}
\label{formula-2}
\int \left(\prod_{m=1}^N d\lambda_m \right) \det\left[ u_j(\lambda_i), 
w_j(\lambda_i) \right]_{1\le i \le N, 1 \le j \le 2N}
= (2N)! \,{\rm pf} \left(
\int d\lambda \left[u_i(\lambda) w_j(\lambda) - u_j(\lambda) w_i(\lambda)\right]
\right)_{1 \le i,j \le 2N}.
\end{equation}
Here, the notation `${\rm pf}$' stands for the pfaffian \cite{M-1989} [its
square is determinant]. The identity Eq. (\ref{formula-2}) enables us to simplify 
Eq. (\ref{os-3}) to
\begin{eqnarray}
\label{os-4a}
{\rm pf} \left(
\int d\lambda f(\lambda)[\varphi_{i-1}(\lambda) \varphi_{j-1}^{\prime}(\lambda)
-\varphi_{j-1}(\lambda) \varphi_{i-1}^{\prime}(\lambda)] 
\right)_{1 \le i,j \le 2N},
\end{eqnarray}
\Rrule
\narrowtext
\noindent
where irrelevant overall constant has been omitted, and the `wave functions' 
$\varphi_j(\lambda)=e^{-2V(\lambda)}P_j(\lambda)$ obeying the orthonormality
relation in the form of Eq. (\ref{wf-2}) have been introduced.
Combining the above results together and properly restoring the normalization, 
we derive the following exact pfaffian 
representation for the averaged product statistics:
\widetext
\Lrule
\begin{eqnarray}
\label{os-6}
\langle F \rangle \equiv 
\frac {{\rm pf} \left(
\int d\lambda f(\lambda)[\varphi_{i-1}(\lambda) \varphi_{j-1}^{\prime}(\lambda)
-\varphi_{j-1}(\lambda) \varphi_{i-1}^{\prime}(\lambda)] 
\right)_{1 \le i,j \le 2N}}
{
{\rm pf} \left(
\int d\lambda [\varphi_{i-1}(\lambda) \varphi_{j-1}^{\prime}(\lambda)
-\varphi_{j-1}(\lambda) \varphi_{i-1}^{\prime}(\lambda)]
\right)_{1 \le i,j \le 2N}
}.
\end{eqnarray}
\Rrule
\narrowtext
\noindent
Equation (\ref{os-6}) is the second preliminary result needed in what follows.

\subsection{Orthogonal symmetry: $\beta=1$}
For $\beta =1$, the pfaffian representation for the averaged product statistics
similar to Eq. (\ref{os-6}) can also be obtained. To avoid unnecessary complications, 
we choose the real symmetric matrix ${\bf H}$ to be of an even dimension, $2N \times 2N$,
so that the upper index in the products in Eqs. (\ref{m-stat}) and (\ref{ms-av}) is equal 
to $2N$. Then, due to the ${\rm O}(2N)$ invariance of the distribution function
$P_{2N}[{\bf H}]$, we encounter the following $2N$--fold integral
\begin{eqnarray}
\label{oso-22}
\langle F \rangle \propto 
\int {\cal D}\hat {\Lambda} e^{-{\rm tr} V(\hat {\Lambda})} 
|\Delta_{2N}(\hat {\Lambda})|
\det [f(\hat {\Lambda})].
\end{eqnarray}
It is useful to reduce Eq. (\ref{oso-22}) to the form \cite{M-1989}
\begin{eqnarray}
\label{or-11}
(2N)!&\int&_{\lambda_1 \le \ldots \le \lambda_{2N}}\left( \prod_{m=1}^{2N} d\lambda_m 
e^{-V(\lambda_m)} f(\lambda_m) \right) \nonumber \\
&\times& \prod_{1 \le i < j \le 2N} (\lambda_j - \lambda_i),
\end{eqnarray}
where the integration runs over the ordered domain $\lambda_1 \le \ldots \le \lambda_{2N}$,
and therefore the modulus of the Vandermonde determinant has been relaxed.
Recalling that the Vandermonde determinant is proportional to 
$\det[P_{j-1}(\lambda_i)]$ with 
$1 \le i,j \le 2N$ and choosing $P_j(\lambda)$ to be the polynomial
orthonormal with respect to the measure $e^{-V(\lambda)}d\lambda$, we are able to 
perform integration in Eq. (\ref{or-11}) by means
of the identity \cite{dB-1955}
\widetext
\Lrule
\begin{equation}
\label{formula-3}
\int_{\lambda_1 \le \ldots \le \lambda_{2N}} \left(\prod_{m=1}^{2N} d\lambda_m\right)
\det[u_j(\lambda_i)]_{1\le i,j \le 2N} = {\rm pf} \left(
\int d\lambda \int d\lambda^{\prime} {\rm sgn}(\lambda^{\prime} -\lambda) u_i(\lambda)
u_j(\lambda^{\prime})
\right)_{1 \le i,j \le 2N},
\end{equation}
where ${\rm sgn}(\lambda) = \lambda/|\lambda|$. The integration results in
\begin{eqnarray}
\label{or-3a}
{\rm pf}\left(
\int d\lambda \int d\lambda^{\prime} \epsilon(\lambda-\lambda^{\prime}) 
f(\lambda) f(\lambda^{\prime}) \varphi_{i-1}(\lambda) \varphi_{j-1}(\lambda^{\prime})
\right)_{1 \le i,j \le 2N},
\end{eqnarray}
\Rrule
\narrowtext
\noindent
where irrelevant overall constant has been suppressed, and the `wave functions'
$\varphi_j(\lambda) = e^{-V(\lambda)/2}P_j(\lambda)$ obeying the orthonormality
relation in the form of Eq. (\ref{wf-2}) have appeared. Also, to meet the standard 
notations of Refs. \cite{M-1991,TW-1998}, we have introduced
\begin{eqnarray}
\label{eps}
\epsilon(\lambda) = \frac{1}{2} {\rm sgn}(\lambda).
\end{eqnarray}
Restoring the proper normalization, one arrives at the third preliminary result:
\widetext
\Lrule
\begin{eqnarray}
\label{os-66}
\langle F \rangle \equiv 
\frac {{\rm pf} \left(
\int \int d\lambda d\lambda^{\prime} 
f(\lambda) f(\lambda^{\prime}) \epsilon(\lambda-\lambda^{\prime})
\varphi_{i-1}(\lambda) \varphi_{j-1}(\lambda^{\prime})
\right)_{1 \le i,j \le 2N}}
{
{\rm pf} \left(
\int \int d\lambda d\lambda^{\prime} \epsilon(\lambda-\lambda^{\prime})
\varphi_{i-1}(\lambda) \varphi_{j-1}(\lambda^{\prime})
\right)_{1 \le i,j \le 2N}
}.
\end{eqnarray}
\Rrule
\narrowtext
\noindent
Notice that Eqs. (\ref{qq}), (\ref{os-6}) and (\ref{os-66}) are exact and refer
to an arbitrary product statistics $F$ that does not affect the convergence of the
integrals under the signs of `${\rm det}$' and `${\rm pf}$'.

\section{Eigenlevel correlations from the replica method}\setcounter{equation}{0}
\subsection{Unitary symmetry: $\beta=2$}

Exact determinantal expression for the averaged replicated partition function
$\langle Z_{\gamma,\gamma^{\prime}}^n (E,E^{\prime}) \rangle$ in the case of the
unitary symmetry is readily obtained from Eq. (\ref{qq}) upon specifying
\begin{eqnarray}
\label{f4b=2}
f(\lambda) = \left[ (E-\lambda)^2 + \gamma^2 \right]^n
\left[ (E^{\prime}-\lambda)^2 + \gamma^{\prime 2} \right]^n ,
\end{eqnarray}
see Eq. (\ref{rpf-intro}). It should be noticed that, contrary to the $\sigma$--model
representation \cite{VZ-1985,KM-1999a,YL-1999,Z-1999} [see, also, Eqs. (\ref{vzi}) and 
(\ref{vzi-det})], our 
determinantal form is especially useful to perform the replica limit $n\rightarrow\pm 0$,
with $+0$ and $-0$ corresponding to the fermionic and the bosonic replicas, respectively.
However, Eq. (\ref{qq}) as it stands, is not convenient for the large--$N$ analysis
of the `density--density' correlation
function. For this reason, we put Eq. (\ref{qq}) into equivalent form
of the functional integral over the scalar Grassmann field
\begin{eqnarray}
\label{grassmann-2}
\psi(\lambda) = \sum_{k=1}^{N} \chi_k \varphi_{k-1}(\lambda)
\end{eqnarray}
with the measure
\begin{eqnarray}
D[{\overline {\psi}},\psi] = \prod_{k=1}^{N} d\chi^*_k d\chi_k ,
\label{measure-2}
\end{eqnarray}
where $\{\chi_k\}$ are $N$ Grassmann complex variables, and ${\overline \psi}$
denotes the complex conjugate. One can verify \cite{Remark-lr4b=2} that the choice Eq. 
(\ref{grassmann-2}) enables us to rewrite Eq. (\ref{qq}) in the form of the functional
integral
\begin{eqnarray}
\label{gi-2}
\langle F \rangle \equiv \int D[{\overline {\psi}},\psi] 
e^{\int \int dx dy {\overline {\psi}}(x)
K_N(x,y)f(y)\psi(y)}
\end{eqnarray}
that solely involves the scalar kernel \cite{M-1991}
\begin{eqnarray}
\label{kernel-2}
K_N(x,y) = \sum_{k=0}^{N-1} \varphi_k(x) \varphi_k(y) = K_N (y,x)
\end{eqnarray} 
obeying the integral relationship
\begin{eqnarray}
\label{2-projector}
\int dz K_N(x,z)K_N(z,y) = K_N(x,y).
\end{eqnarray}
In turn, the universal properties of the scalar kernel $K_N(x,y)$ in the large--$N$ limit are well 
studied \cite{BZ-1993,KF-1999}.

Substituting Eq. (\ref{f4b=2}) into Eq. (\ref{gi-2}), differentiating
the latter with respect to $\gamma$ and $\gamma^{\prime}$ and
computing the limits $n\rightarrow \pm 0$ and 
$\gamma, \gamma^{\prime} \rightarrow 0$ one gets access to the `density--density' correlation 
function in accordance with Eq. (\ref{limit-dos-dos}). Since we are interested in the
correlations in the energy range where $E-E^{\prime}$ may take the values of order
of the mean level spacing $\Delta_N$, we introduce the scaled variables $S=E/\Delta_N$ and
$S^{\prime} = E^{\prime}/\Delta_N$, and restrict ourselves to the vicinity of
the center of the eigenvalue support, where $S,S^{\prime} \sim {\cal O}(N^0)$. In
the large--$N$ limit, the rescaled Eq. (\ref{limit-dos-dos}) is then reduced to
\begin{eqnarray}
\langle \rho(S) \rho(S^{\prime})\rangle &=& \Delta_N ^2 
\lim_{\gamma,\gamma^{\prime}\rightarrow 0} \lim_{n\rightarrow \pm 0}
\frac{1}{(2\pi n)^2} \nonumber \\
&\times& \partial_\gamma \partial_{\gamma^{\prime}}
\langle Z_{\gamma,\gamma^{\prime}}^n (S\Delta_N, S^{\prime}\Delta_N)\rangle .
\label{roro-2}
\end{eqnarray}
Straightforward calculations based on Eqs. (\ref{roro-2}) and (\ref{gi-2}) with
$f(\lambda)$ given by Eq. (\ref{f4b=2}) yield
\widetext
\Lrule
\begin{eqnarray}
\label{uu-2}
\langle \rho(S) \rho(S^{\prime}) \rangle &=& \delta(S-S^{\prime}) \Delta_N^2
\int d\sigma K_N(\sigma \Delta_N, S\Delta_N) \langle \langle {\overline \psi}
(\sigma \Delta_N) \psi(S\Delta_N) \rangle\rangle \nonumber \\
&+& \Delta_N ^4 \int \int d\sigma d\sigma^{\prime} K_N(\sigma \Delta_N, S\Delta_N)
\langle \langle
{\overline \psi}(\sigma \Delta_N) \psi(S\Delta_N) 
{\overline \psi}(\sigma^{\prime}\Delta_N) \psi(S^{\prime}\Delta_N)
\rangle \rangle
K_N (\sigma^{\prime}\Delta_N, S^{\prime} \Delta_N) ,
\end{eqnarray}
where the double angular brackets $\langle\langle \ldots \rangle \rangle$ stand for 
the Gaussian average
\begin{eqnarray}
\langle\langle \ldots \rangle\rangle = \int D[{\overline \psi},\psi] (\ldots)
e^{\Delta_N ^2 \int\int d\sigma d\sigma^{\prime} {\overline \psi}(\sigma \Delta_N)
K_N(\sigma \Delta_N, \sigma^{\prime} \Delta_N) \psi (\sigma^{\prime}\Delta_N)}.
\label{gauss-aver}
\end{eqnarray}
In Eq. (\ref{uu-2}), $\langle\langle {\overline \psi}\psi \rangle \rangle$ and 
$\langle\langle {\overline \psi}\psi {\overline \psi}\psi \rangle \rangle$ 
may be computed via the Wick theorem \cite{Remark-Wick}:
\begin{mathletters}
\label{wick-2}
\begin{eqnarray}
\langle\langle {\overline \psi}(\sigma_1\Delta_N) \psi(\sigma_2\Delta_N) \rangle
\rangle
&=& K_N(\sigma_1\Delta_N,\sigma_2\Delta_N), 
\label{wick-2a} \\
\langle \langle
{\overline \psi}(\sigma_1\Delta_N) \psi(\sigma_2\Delta_N)
{\overline \psi}(\sigma_3\Delta_N) \psi(\sigma_4\Delta_N)
\rangle \rangle 
&=& K_N(\sigma_1\Delta_N,\sigma_2\Delta_N)K_N(\sigma_3\Delta_N,\sigma_4\Delta_N) 
\nonumber \\
&-& K_N(\sigma_1\Delta_N,\sigma_4\Delta_N)
K_N(\sigma_3\Delta_N,\sigma_2\Delta_N). \label{wick-2b}
\end{eqnarray}
\end{mathletters}
\Rrule
\narrowtext
\noindent

As far as we are concerned with the spectrum bulk, one has
$\Delta_{N}K_N(S \Delta_N, S^{\prime} \Delta_N) = K(s)$ with 
\begin{eqnarray}
K(s) = \frac{\sin(\pi s)}{\pi s}, \label{sin-ker}
\end{eqnarray}
$s=|S-S^{\prime}|$, by the universality arguments \cite{BZ-1993,KF-1999,M-1991}. 
This equation holds for the strong confinement potential \cite{Remark1} only.

Combining now Eqs. (\ref{uu-2}), (\ref{wick-2}) and (\ref{sin-ker}), we come down to
\begin{eqnarray}
\label{uu-3}
\langle \rho(S) \rho(S^{\prime}) \rangle =
1 - K^2(s)
+ \delta(s)
\end{eqnarray}
which is clearly identical to Eq. (\ref{unit}). This is the central result of the 
subsection.

It must be stressed that, by derivation, Eq. (\ref{uu-3}) is valid in the bulk 
of the spectrum {\it without} the restriction $s \gg 1$ inherent
to the approximate calculational schemes \cite{KM-1999a,YL-1999}. The self-correlation of 
energy levels is also reproduced.

\subsection{Symplectic symmetry: $\beta=4$}
For technical reasons dictated by the pfaffian structure of Eq. (\ref{os-6}) derived for
$\beta=4$, it is useful to deal with 
$\langle Z_{\gamma,\gamma^{\prime}}^n (E,E^{\prime})\rangle ^2$
instead of $\langle Z_{\gamma,\gamma^{\prime}}^n (E,E^{\prime})\rangle$. In such a case, 
the `density--density' correlation function is seen to be generated by the replica limit
\widetext
\Lrule
\begin{eqnarray}
\label{rl4b=4}
\langle \rho_N(E) \rho_N(E^{\prime})\rangle = \frac{1}{2} 
\lim_{\gamma,\gamma^{\prime}\rightarrow  0} \lim_{n \rightarrow \pm 0}
\frac{1}{(2\pi n)^2} \left[
\partial_\gamma \partial_{\gamma^{\prime}} \langle Z_{\gamma,\gamma^{\prime}}^n
(E,E^{\prime})\rangle ^2 - \frac{1}{2} \partial_\gamma 
\langle Z_{\gamma,\gamma^{\prime}}^n
(E,E^{\prime})\rangle ^2
\partial_{\gamma^{\prime}}
\langle Z_{\gamma,\gamma^{\prime}}^n
(E,E^{\prime})\rangle ^2
\right],
\end{eqnarray}
\Rrule
\narrowtext
\noindent
with $\langle Z_{\gamma,\gamma^{\prime}}^n(E,E^{\prime})\rangle ^2$ being the ratio of two 
determinants, Eq. (\ref{os-6});
the function $f(\lambda)$ is given by Eq. (\ref{f4b=2}).

To facilitate the large--$N$ analysis, we rewrite the squared Eq. (\ref{os-6})
in the equivalent form of the ratio of two Grassmann functional integrals:
\begin{eqnarray}
\label{gi-4}
\langle F \rangle ^2 \equiv \frac
{
\int D[{\overline {\Psi}},\Psi] e^{\int \int dx dy {\overline {\Psi}}(x) {\hat K}_N(x,y)
f(y) \Psi(y)}
}
{
\int D[{\overline {\Psi}},\Psi] e^{\int \int dx dy {\overline {\Psi}}(x) {\hat K}_N(x,y)
\Psi(y)}
}.
\end{eqnarray}
Here, $\Psi(x)$ is the two-component Grassmann field
\begin{eqnarray}
\label{gf-4}
\Psi(x) = \sum_{k=1}^{2N} \chi_k \phi_{k-1}(x) = \sum_{k=1}^{2N} \chi_k \left( 
\begin{array}{c}
\varphi_{k-1}^{\prime}(x) \\ 
\varphi_{k-1}(x)
\end{array}
\right),
\end{eqnarray}
and ${\overline {\Psi}}(x) = \Psi^{\dagger}(x)L$, with
\begin{eqnarray}
\label{l-m}
L = \left( 
\begin{array}{cc}
0 & -1 \\ 
1 & 0
\end{array}
\right). 
\end{eqnarray}
Also, the integration measure is
\begin{eqnarray}
D[{\overline {\Psi}},\Psi] = \prod_{k=1}^{2N} d\chi^*_k
d\chi_k.
\label{im-4}
\end{eqnarray}

The choice Eq. (\ref{gf-4}) naturally leads to appearance of the $2 \times 2$ matrix 
kernel 
\begin{eqnarray}
\label{k-4}
{\hat K}_N (x,y) 
= \left( 
\begin{array}{cc}
K_N^{11}(x,y) & 
K_N^{12}(x,y) \\ 
K_N^{21}(x,y) & 
K_N^{22}(x,y)
\end{array}
\right)
\end{eqnarray}
which is a close analog of the scalar kernel $K_N(x,y)$, Eq. (\ref{kernel-2}), arising 
in random matrix ensembles with ${\rm U}(N)$ symmetry. In terms of the 
wave functions $\varphi_k (x)$ of Eq. (\ref{os-6}), the $(\alpha,\beta)$ entry of the 
matrix kernel is given by
\begin{equation}
\label{k-44}
{\hat K}_N^{\alpha \beta} (x,y) 
= (-1)^{\beta-1}\sum_{j,k=1}^{2N} \varphi_{j-1}^{(2-\alpha)}(x) \mu_{jk} 
\varphi_{k-1}^{(\beta-1)}(y),
\end{equation}
[$\alpha,\beta =1,2$], so that the integral identity
\begin{eqnarray}
\label{4-projector}
\int dz {\hat K}_N(x,z) {\hat K}_N(z,y) = {\hat K}_N(x,y)
\end{eqnarray}
holds. Here, $\varphi_k ^{(\alpha)}(x)$ denotes the derivative of
$\alpha$--th order with respect to $x$, and $\mu_{jk} = (M^{-1})_{jk}$, where $2N\times 2N$
antisymmetric matrix $M$ has the entries
\begin{eqnarray}
\label{m-matrix-4}
M_{jk} = \int d\lambda 
[
\varphi_{j-1}(\lambda) \varphi_{k-1}^{\prime}(\lambda)
-\varphi_{k-1}(\lambda) \varphi_{j-1}^{\prime}(\lambda)
],
\end{eqnarray}
$j,k \in (1,\ldots,2N)$. 
Equation (\ref{k-44}) coincides with the one found for the first time by Tracy and 
Widom \cite{TW-1998}. As is the case of ${\rm U}(N)$ symmetry, the large--$N$ behavior
of the matrix kernel ${\hat K}_N$ is known in detail \cite{M-1991,W-1999,SV-1998}.

The representation Eq. (\ref{gi-4}) is a bit less trivial as compared to 
Eq. (\ref{gi-2}). To verify the former, one should observe that 
$\phi_k(x)$ in Eq. (\ref{gf-4}) is the right eigenvector of the matrix kernel 
${\hat K}_N(x,y)$:
\begin{eqnarray}
\label{rev-4}
\int dy {\hat K}_N(x,y) \phi_k(y) = \phi_k(x),
\end{eqnarray}
while ${\overline \phi}_k(x)=\phi_k^{\dagger}(x)L$
is its left eigenvector:
\begin{eqnarray}
\label{lev-4}
\int dx {\overline \phi}_k(x) {\hat K}_N(x,y) = {\overline \phi}_k(y),
\end{eqnarray}
with $k \in (0,\ldots,2N-1)$.

In what follows we are interested in the `density--density' correlation function in rescaled
energy variables $S=E/\Delta_N$ and $S^{\prime}=E^{\prime}/\Delta_N$, when the
large--$N$ limit of Eq. (\ref{rl4b=4}) transforms to
\widetext
\Lrule
\begin{eqnarray}
\label{rl4b=44}
\langle \rho(S) \rho(S^{\prime})\rangle &=& \frac{\Delta_N ^2}{2} 
\lim_{\gamma,\gamma^{\prime}\rightarrow  0} \lim_{n \rightarrow \pm 0}
\frac{1}{(2\pi n)^2} \nonumber \\
&\times& \left[
\partial_\gamma \partial_{\gamma^{\prime}} \langle Z_{\gamma,\gamma^{\prime}}^n
(S\Delta_N,S^{\prime}\Delta_N)\rangle ^2 - \frac{1}{2} \partial_\gamma 
\langle Z_{\gamma,\gamma^{\prime}}^n
(S\Delta_N,S^{\prime}\Delta_N)\rangle ^2
\partial_{\gamma^{\prime}}
\langle Z_{\gamma,\gamma^{\prime}}^n
(S\Delta_N,S^{\prime}\Delta_N)\rangle ^2
\right].
\end{eqnarray}

Substituting Eq. (\ref{gi-4}) with $f(\lambda)$ given by Eq. (\ref{f4b=2}) into
Eq. (\ref{rl4b=44}) leads us to the following representation:
\begin{eqnarray}
\label{tlcf4b=4}
\langle \rho (S) \rho (S^{\prime}) \rangle &=& \frac{1}{2}
\left[
\delta(S-S^{\prime}) \Delta_N^2 \int d\sigma \sum_{\alpha,\beta =1}^2 
{\hat K}_N^{\alpha \beta}(\sigma\Delta_N,S\Delta_N) 
\langle\langle {\overline \Psi}^{\alpha}(\sigma\Delta_N) \Psi^{\beta}(S\Delta_N) 
\rangle\rangle \right. \nonumber \\
&+&
\Delta_N^4 \int\int d\sigma d\sigma^{\prime} \sum_{\alpha,\beta=1}^2
\sum_{\gamma,\delta =1}^2
{\hat K}_N^{\alpha \beta}(\sigma\Delta_N,S\Delta_N) 
{\hat K}_N^{\gamma \delta}(\sigma^{\prime}\Delta_N,S^{\prime}\Delta_N) \nonumber
\\
&\times&
\left.
\left(
\langle \langle 
{\overline \Psi}^{\alpha}(\sigma\Delta_N) \Psi^{\beta}(S\Delta_N) 
{\overline \Psi}^{\gamma}(\sigma^{\prime}\Delta_N) \Psi^{\delta}(S^{\prime}\Delta_N)
\rangle \rangle \right. \right. \nonumber \\
&-& \left. \left.
\frac{1}{2} 
\langle\langle {\overline \Psi}^{\alpha}(\sigma\Delta_N)
\Psi^{\beta}(S\Delta_N)\rangle\rangle
\langle\langle {\overline \Psi}^{\gamma}(\sigma^{\prime}\Delta_N)
\Psi^{\delta}(S^{\prime}\Delta_N)\rangle\rangle
\right)
\right].
\end{eqnarray}
Here, the double angular brackets $\langle\langle \ldots \rangle\rangle$ denote the
Gaussian average
\begin{equation}
\label{aver-4}
\langle\langle \ldots \rangle\rangle = \frac{\int D[{\overline \Psi},\Psi]
(\ldots) e^{
{\Delta_N^2 \int \int d\sigma d\sigma^{\prime} {\overline {\Psi}}(\sigma\Delta_N) 
{\hat K}_N(\sigma\Delta_N,\sigma^{\prime}\Delta_N) \Psi(\sigma^{\prime}\Delta_N)}
}
}
{\int D[{\overline \Psi},\Psi] e^{
{\Delta_N^2 \int \int d\sigma d\sigma^{\prime} {\overline {\Psi}}(\sigma\Delta_N) 
{\hat K}_N(\sigma\Delta_N,\sigma^{\prime}\Delta_N) \Psi(\sigma^{\prime}\Delta_N)}
}
}.
\end{equation}
The averages of the type $\langle\langle {\overline \Psi}^{\alpha} \Psi^{\beta} \rangle\rangle$
and $\langle\langle {\overline \Psi}^{\alpha} \Psi^{\beta} {\overline \Psi}^{\gamma} \Psi^{\delta} \rangle\rangle$
in Eq. (\ref{tlcf4b=4}) are evaluated by means of the Wick theorem [see Appendix
for details]; below we display the result:
\begin{mathletters}
\label{gauss-4}
\begin{eqnarray}
\langle \langle
{\overline \Psi}^{\alpha}(\sigma_1\Delta_N) \Psi^{\beta} (\sigma_2\Delta_N ) 
\rangle \rangle &=&
{\hat K}_N^{\beta \alpha} (\sigma_2\Delta_N,
\sigma_1\Delta_N), \label{gauss-4a} \\
\langle\langle 
{\overline \Psi}^{\alpha}(\sigma_1\Delta_N) \Psi^{\beta}(\sigma_2\Delta_N )
{\overline \Psi}^{\gamma}(\sigma_3\Delta_N) \Psi^{\delta}(\sigma_4\Delta_N )  
\rangle \rangle
&=& {\hat K}_N^{\beta \alpha} (\sigma_2\Delta_N, \sigma_1\Delta_N)
{\hat K}_N^{\delta \gamma} (\sigma_4\Delta_N,\sigma_3\Delta_N) \nonumber \\
&-& {\hat K}_N^{\delta \alpha}(\sigma_4\Delta_N, \sigma_1\Delta_N)
{\hat K}_N^{\beta \gamma} (\sigma_2\Delta_N, \sigma_3\Delta_N) 
\label{gauss-4b}.
\end{eqnarray}
\end{mathletters}
Inserting Eqs. (\ref{gauss-4}) into Eq. (\ref{tlcf4b=4}), and making use
of Eq. (\ref{4-projector}), one obtains:
\begin{eqnarray}
\label{rr-4}
\langle \rho(S)\rho(S^{\prime}) \rangle &=& \frac{1}{2}\left[
\delta(S-S^{\prime}) \Delta_N {\rm tr} [{\hat K}_N(S\Delta_N,S\Delta_N)] \right. 
\nonumber \\
&+& \left. \Delta_N^2\left(
\frac{1}{2} {\rm tr}[{\hat K}_N(S\Delta_N, S\Delta_N)]
{\rm tr}[{\hat K}_N(S^{\prime}\Delta_N,S^{\prime}\Delta_N)]
- {\rm tr} [{\hat K}_N(S\Delta_N,S^{\prime}\Delta_N) 
{\hat K}_N(S^{\prime}\Delta_N,S\Delta_N)]
\right)
\right].
\end{eqnarray}
\Rrule
\narrowtext
\noindent
Here, the trace `${\rm tr}$' acts in the $2\times 2$ space of the matrix kernel ${\hat K}_N$.

In the large--$N$ limit, the kernel ${\hat K}_N(S\Delta_N,S^{\prime}\Delta_N)$ taken 
in the bulk of the spectrum [which is of our primary interest] obeys the universal
laws \cite{W-1999,SV-1998,M-1991} 
\begin{mathletters}
\label{univ-4}
\begin{eqnarray}
\Delta_N {\hat K}_N^{11} (S\Delta_N,S^{\prime}\Delta_N) &=&
\Delta_N {\hat K}_N^{22} (S\Delta_N,S^{\prime}\Delta_N) \nonumber \\
&=& K(s), \label{univ-4a} \\
\Delta_N^2 {\hat K}_N^{12} (S\Delta_N,S^{\prime}\Delta_N) &=& K^{\prime}(s), \label{univ-4b} \\
{\hat K}_N^{21} (S\Delta_N,S^{\prime}\Delta_N) &=& \int_0^{s} dt K(t), \label{univ-4c}
\end{eqnarray}
\end{mathletters}
provided the confinement potential is strong \cite{Remark1}. Here, $s=|S-S^\prime|$, and
\begin{equation}
\label{kkk}
K(s) = \frac{\sin(2\pi s)}{2\pi s}.
\end{equation}

Combining Eqs. (\ref{rr-4}), (\ref{univ-4}) and (\ref{kkk}), we come down to
\begin{equation}
\label{final-4}
\langle \rho(S) \rho(S^{\prime})\rangle = 1 - K^2(s) + \frac{dK(s)}{ds}\int_0^s dt K(t)
+\delta(s).
\end{equation}
Equation (\ref{final-4}), derived from the replicas, is seen to identically 
coincide with Eq. 
(\ref{sympl}) for arbitrary $s$. It represents the main result of the subsection.

\subsection{Orthogonal symmetry: $\beta=1$}
As is the case of the symplectic symmetry, it is useful to deal with 
$\langle Z_{\gamma,\gamma^{\prime}}^n(E,E^{\prime})\rangle ^2$, so that the 
`density--density' correlation function is defined by the replica
limit Eq. (\ref{rl4b=4}) with $N$ being replaced by $2N$, and with 
$\langle Z_{\gamma,\gamma^{\prime}}^n(E,E^{\prime})\rangle ^2$
given by the ratio of two determinants as follows from Eq. (\ref{os-66});
the function $f(\lambda)$ is specified by Eq. (\ref{f4b=2}).

In order to simplify the large--$N$ analysis, we have to express 
$\langle Z_{\gamma,\gamma^{\prime}}^n(E,E^{\prime})\rangle ^2$
in terms of the Grassmann functional integrals. One can verify that
the squared averaged partition function admits the representation
\widetext
\Lrule
\begin{eqnarray}
\label{gi-1}
\langle F \rangle ^2 \equiv \frac
{
\int D[{\overline \Psi},\Psi] e^{\int\int dx dy {\overline \Psi}(x)f(x)
[{\hat K}_{2N}(x,y) - {\hat Q}_{2N}(x,y)]f(y)\Psi(y)}
}
{
\int D[{\overline \Psi},\Psi] e^{\int\int dx dy {\overline \Psi}(x)
[{\hat K}_{2N}(x,y) - {\hat Q}_{2N}(x,y)]\Psi(y)}
}.
\end{eqnarray}
\Rrule
\narrowtext
\noindent
Here, the integration measure $D[{\overline \Psi},\Psi]$ is given by Eq. (\ref{im-4}).
The two-component Grassmann field $\Psi$ is defined as
\begin{eqnarray}
\label{tcgf-1}
\Psi(x) = \sum_{k=1}^{2N} \chi_k \phi_{k-1}(x) = \sum_{k=1}^{2N}
\chi_k 
\left( 
\begin{array}{c}
\varphi_{k-1}(x) \\ 
\left[ {\hat \epsilon} \varphi_{k-1} \right](x)
\end{array}
\right),
\end{eqnarray}
with ${\overline \Psi}(x) = \Psi^{\dagger}(x)L$; $2\times 2$ matrix $L$ is given by 
Eq. (\ref{l-m}).
Also, ${\hat \epsilon}$ denotes the integral operator associated with Eq. (\ref{eps});
it acts on some function
$g(x)$ in accordance with the rule
\begin{eqnarray}
[{\hat \epsilon} g](x) = \int dy \epsilon (x-y) g(y)
\end{eqnarray}
provided the integral is convergent.

The representation Eq. (\ref{gi-1}) involves the $2\times 2$ matrices
${\hat Q}_{2N}$ and ${\hat K}_{2N}$ which are related to each other
as
\begin{eqnarray}
\label{ker-1}
{\hat Q}_{2N}(x,y) = {\hat K}_{2N}(x,y) + 
\left( 
\begin{array}{cc}
0 & 
0 \\ 
\epsilon(x-y) & 
0
\end{array}
\right)
\end{eqnarray}
with
\begin{equation}
\label{Q-eq}
{\hat Q}_{2N}^{\alpha \beta}(x,y) = (-1)^\beta \sum_{j,k=1}^{2N}[{\hat \epsilon}^{\alpha-1}\varphi_{j-1}](x)
\mu_{jk}[{\hat \epsilon}^{2-\beta}\varphi_{k-1}](y),
\end{equation}
$[\alpha,\beta=1,2]$. Here, the matrix $\mu_{jk} = (M^{-1})_{jk}$, where $2N\times 2N$
antisymmetric matrix $M$ has the entries
\begin{eqnarray}
\label{M-1}
M_{jk} = \int \int dx dy \varphi_{j-1}(x)\epsilon(x-y)\varphi_{k-1}(y),
\end{eqnarray}
$j,k\in(1,\ldots,2N)$.

Let us point out that the matrix ${\hat K}_{2N}(x,y)$ is nothing but the 
$2\times 2$ matrix kernel \cite{TW-1998}
describing energy level correlations in random matrix ensembles with orthogonal 
symmetry. However, contrary to Eq. (\ref{4-projector}), the kernel ${\hat K}_{2N}(x,y)$
obeys the more complicated integral identity
\begin{eqnarray}
\label{projector-1}
\int dz {\hat K}_{2N}(x,z){\hat K}_{2N}(z,y) &=& {\hat K}_{2N}(x,y) \nonumber \\
&-& \frac{1}{2}
[{\hat K}_{2N}(x,y),\Lambda],
\end{eqnarray}
where $[\ldots,\ldots]$ stands for commutator, and the matrix
$\Lambda= {\rm diag}(+1,-1)$. The auxiliary matrix ${\hat Q}_{2N}(x,y)$ 
satisfies
\begin{eqnarray}
\label{Q-projector}
\int dz {\hat Q}_{2N}(x,z){\hat Q}_{2N}(z,y) = 2 {\hat Q}_{2N}(x,y).
\end{eqnarray}

An advantage of the representation Eq. (\ref{gi-1}) is in the fact that the
Gaussian averages of the type $\langle\langle{\overline \Psi}^{\alpha} \Psi^\beta\rangle\rangle$ and 
$\langle\langle{\overline \Psi}^\alpha \Psi^\beta {\overline \Psi}^\gamma \Psi^\delta\rangle\rangle$ arising
when implementing the replica limit may easily be expressed in terms of the
large--$N$ matrix kernel ${\hat K}_{2N}$; the universal properties of the latter 
for the strong level confinement \cite{Remark1} are 
well studied \cite{W-1999,SV-1998,M-1991}.

Restricting ourselves to the spectrum bulk, the `density--density' correlation
function in rescaled variables $S=E/\Delta_{2N}$ and $S^{\prime}=E^{\prime}/\Delta_{2N}$
is given by the replica limit Eq. (\ref{rl4b=44}). Substituting there Eq. (\ref{gi-1})
with $f(\lambda)$ specified by Eq. (\ref{f4b=2}) and taking into account Eq. (\ref{ker-1}),
we obtain
\widetext
\Lrule
\begin{eqnarray}
\label{dd-1}
\langle \rho(S) \rho(S^{\prime}) \rangle &=& \frac{\Delta_{2N}^2}{2}\left\{
\epsilon(S-S^{\prime}) \left[ \langle\langle {\overline \Psi}^2(S^{\prime}\Delta_{2N})
\Psi^1 (S\Delta_{2N})\rangle\rangle - \langle\langle {\overline \Psi}^2(S\Delta_{2N})
\Psi^1 (S^{\prime}\Delta_{2N}) 
\rangle\rangle
\right]
\right. \nonumber \\
&+& \left. \delta(S-S^{\prime}) \int d\sigma \epsilon(S-\sigma)\left[
\langle\langle {\overline \Psi}^2(\sigma\Delta_{2N})
\Psi^1 (S\Delta_{2N})\rangle\rangle - \langle\langle {\overline \Psi}^2(S\Delta_{2N})
\Psi^1 (\sigma\Delta_{2N}) 
\rangle\rangle
\right] \right. \nonumber \\
&+& \left.
\Delta_{2N}^2\int \int d\sigma d\sigma^{\prime} \epsilon(S^{\prime}-\sigma)
\epsilon(S-\sigma^{\prime}) \left[
\langle\langle {\overline \Psi}^2(S^{\prime}\Delta_{2N})\Psi^1(\sigma\Delta_{2N})
{\overline \Psi}^2(S\Delta_{2N})\Psi^1(\sigma^{\prime}\Delta_{2N}) \rangle\rangle
\right. \right. \nonumber \\
&+& \left.\left. 
\langle\langle {\overline \Psi}^2(\sigma\Delta_{2N})\Psi^1(S^{\prime}\Delta_{2N})
{\overline \Psi}^2(\sigma^{\prime}\Delta_{2N})\Psi^1(S\Delta_{2N}) \rangle\rangle
\right. \right. \nonumber \\
&-& \left. \left.
\langle\langle {\overline \Psi}^2(S^{\prime}\Delta_{2N})\Psi^1(\sigma\Delta_{2N})
{\overline \Psi}^2(\sigma^{\prime}\Delta_{2N})\Psi^1(S\Delta_{2N}) \rangle\rangle
\right. \right. \nonumber \\
&-& \left. \left.
\langle\langle {\overline \Psi}^2(\sigma\Delta_{2N})\Psi^1(S^{\prime}\Delta_{2N})
{\overline \Psi}^2(S\Delta_{2N})\Psi^1(\sigma^{\prime}\Delta_{2N}) \rangle\rangle
\right. \right. \nonumber \\
&-& \left. \left. \frac{1}{2} \left( \langle\langle {\overline \Psi}^2(S^{\prime}\Delta_{2N})
\Psi^1(\sigma\Delta_{2N}) \rangle\rangle
- \langle\langle
{\overline \Psi}^2(\sigma\Delta_{2N})\Psi^1(S^{\prime}\Delta_{2N})
\rangle\rangle
\right) \right. \right. \nonumber \\
&\times& \left. \left.
\left(
\langle\langle {\overline \Psi}^2(S\Delta_{2N})
\Psi^1(\sigma^{\prime}\Delta_{2N}) \rangle\rangle
- \langle\langle
{\overline \Psi}^2(\sigma^{\prime}\Delta_{2N})\Psi^1(S\Delta_{2N})
\rangle\rangle
\right)
\right]
\right\}.
\end{eqnarray}

Here, the notation $\langle\langle\ldots\rangle\rangle$ stands for the
Gaussian average
\begin{eqnarray}
\label{gauss-1}
\langle\langle\ldots\rangle\rangle = 
\frac
{
\int D[{\overline \Psi},\Psi] (\ldots) e^{\Delta_{2N}^2\int\int d\sigma d\sigma^{\prime} 
{\overline \Psi}(\sigma\Delta_{2N})
[{\hat K}_{2N}(\sigma\Delta_{2N},\sigma^{\prime}\Delta_{2N}) - 
{\hat Q}_{2N}(\sigma\Delta_{2N},\sigma^{\prime}\Delta_{2N})]\Psi(\sigma^{\prime}\Delta_{2N})}
}
{
\int D[{\overline \Psi},\Psi] e^{\Delta_{2N}^2\int\int d\sigma d\sigma^{\prime} 
{\overline \Psi}(\sigma\Delta_{2N})
[{\hat K}_{2N}(\sigma\Delta_{2N},\sigma^{\prime}\Delta_{2N}) - 
{\hat Q}_{2N}(\sigma\Delta_{2N},\sigma^{\prime}\Delta_{2N})]\Psi(\sigma^{\prime}\Delta_{2N})}
}.
\end{eqnarray}

For the averages $\langle\langle \ldots \rangle\rangle$ the Wick
theorem holds [see Appendix for details], so that
\begin{mathletters}
\label{wick-1}
\begin{eqnarray}
\langle\langle {\overline \Psi}^\alpha (\sigma_1\Delta_{2N}) \Psi^\beta(\sigma_2\Delta_{2N})
\rangle\rangle &=& - {\hat Q}_{2N}^{\beta\alpha}(\sigma_2\Delta_{2N},\sigma_1\Delta_{2N}),
\label{w-1a} \\
\langle\langle {\overline \Psi}^\alpha (\sigma_1\Delta_{2N}) \Psi^\beta(\sigma_2\Delta_{2N})
{\overline \Psi}^\gamma (\sigma_3\Delta_{2N}) \Psi^\delta (\sigma_4\Delta_{2N})
\rangle\rangle &=& {\hat Q}_{2N}^{\beta\alpha}(\sigma_2\Delta_{2N},\sigma_1\Delta_{2N})
{\hat Q}_{2N}^{\delta\gamma}(\sigma_4\Delta_{2N},\sigma_3\Delta_{2N}) \nonumber \\
&-& 
{\hat Q}_{2N}^{\delta\alpha}(\sigma_4\Delta_{2N},\sigma_1\Delta_{2N})
{\hat Q}_{2N}^{\beta\gamma}(\sigma_2\Delta_{2N},\sigma_3\Delta_{2N}). 
\label{w-1b}
\end{eqnarray}
\end{mathletters}
Inserting Eqs. (\ref{wick-1}) into Eq. (\ref{dd-1}) and taking into account
Eqs. (\ref{ker-1}) and (\ref{Q-eq}), one derives after some algebra:
\begin{eqnarray}
\label{dd-11}
\langle \rho(S)\rho(S^{\prime}) \rangle &=& 
\delta(S-S^{\prime})\Delta_{2N}{\hat K}_{2N}^{11}(S\Delta_{2N},S\Delta_{2N})
+ \Delta_{2N}^2 \left[ 
{\hat K}_{2N}^{11}(S\Delta_{2N},S\Delta_{2N})
{\hat K}_{2N}^{11}(S^{\prime}\Delta_{2N},S^{\prime}\Delta_{2N}) 
\right.  \nonumber \\
&-& \left.
{\hat K}_{2N}^{11}(S\Delta_{2N},S^{\prime}\Delta_{2N})
{\hat K}_{2N}^{22}(S\Delta_{2N},S^{\prime}\Delta_{2N})
+
{\hat K}_{2N}^{12}(S\Delta_{2N},S^{\prime}\Delta_{2N})
{\hat K}_{2N}^{21}(S\Delta_{2N},S^{\prime}\Delta_{2N})
\right].
\end{eqnarray} 
\Rrule
\narrowtext
\noindent

In the large--$N$ limit, the kernel ${\hat K}_{2N}(S\Delta_{2N},S^{\prime}\Delta_{2N})$ 
taken in the bulk of the spectrum is known to obey the universal
laws \cite{W-1999,SV-1998,M-1991} 
\begin{mathletters}
\label{K-un-1}
\begin{eqnarray}
\Delta_{2N} {\hat K}_{2N}^{11} (S\Delta_{2N},S^{\prime}\Delta_{2N}) &=&
\Delta_{2N} {\hat K}_{2N}^{22} (S\Delta_{2N},S^{\prime}\Delta_{2N}) \nonumber \\
&=& K(s), \label{K-un-1a} \\
\Delta_{2N}^2 {\hat K}_{2N}^{12} (S\Delta_{2N},S^{\prime}\Delta_{2N}) &=& K^{\prime}(s), 
\label{K-un-1b} \\
{\hat K}_{2N}^{21} (S\Delta_{2N},S^{\prime}\Delta_{2N}) &=& -\int_s^{+\infty} dt K(t), 
\label{K-un-1c}
\end{eqnarray}
\end{mathletters}
provided the confinement potential is strong \cite{Remark1}. Here, $s=|S-S^\prime|$, 
and 
\begin{equation}
\label{k-1}
K(s) = \frac{\sin(\pi s)}{\pi s}.
\end{equation}

Taking into account Eqs. (\ref{K-un-1}) and (\ref{k-1}), one concludes that
Eq. (\ref{dd-11}) [which is valid for arbitrary $s$] is equivalent to
\begin{equation}
\label{answer-1}
\langle \rho(S) \rho(S^{\prime}) \rangle = 1-K^2(s) -\frac{dK(s)}{ds}\int_s^{+\infty}
dt K(t) + \delta(s),
\end{equation}
that, in turn, is identical to Eq. (\ref{orth}) as one could already expect.

\section{Conclusions}\setcounter{equation}{0}
The computation presented above has been inspired by the recent paper
\cite{KM-1999a} by Kamenev and M\'ezard, followed by the further developments 
\cite{KM-1999b,YL-1999,Z-1999}, where the problem of the fermionic replica method has been
revisited. The main outcome of Refs. \cite{KM-1999a,KM-1999b,YL-1999} consists of
assertion that the replica method in its fermionic $\sigma$--model elaboration,
which relies on a selected set of causal saddle points \cite{Z-1999}, is capable of 
reproducing the nonperturbative results for the spectral statistics, both in Gaussian 
invariant ensembles \cite{KM-1999a,YL-1999} in the thermodynamic limit, and in 
$d$--dimensional disordered conductors with finite Thouless conductance, $g \gg 1$,
albeit in the large--$s$ domain, $s \gg 1$.

In this paper we examined the replica method from a different point of view, considering
the fermionic and the bosonic replicas on an equal footing. The crucial difference between 
the previous studies and the present consideration resides in using an alternative 
[to the $\sigma$--model approach] technique to average the replicated partition function
over the ensemble of random matrices for all three Dyson symmetry classes. This technique
generates such representations for the spectral observables which are especially suitable
for taking the replica limit $n\rightarrow \pm 0$ in a straightforward way. Performing
exact [large--$N$] transformations at each stage, we have derived the nonperturbative 
random--matrix--theory fluctuation formulas for the spectral 
correlation functions in {\it arbitrary} spectrum range, including the effect of
self-correlation of the energy levels showing up in the $\delta$--functional contribution
in Eqs. (\ref{uu-3}), (\ref{final-4}) and (\ref{answer-1}). Let us notice that the 
$\delta$--functional contribution is not an academic issue: It is
this $\delta$--function whose spreading \cite{KZ-1992} [in the presence of an external 
perturbation] determines such important characteristics of disordered system as
distributions of level velocities \cite{THSA-1994} and level curvatures \cite{O-1994}.

Reproducing of exact fluctuation formulas for invariant random matrix ensembles
at $\beta=1,2$ and $4$ through the replica method comprises the main outcome of our study.
It explicitly demonstrates that, at least in the context of invariant random matrix models, the replica 
method {\it itself} is free from internal subtleties and thus completely legitimate,
provided one is able to control the analytic properties of the replicated partition
function Eq. (\ref{rpf-intro}) as a function of the replica parameter $n$ after the 
averaging over realizations of the random matrix Hamiltonian ${\bf H}$. The 
availability of such a control in the $\sigma$--model--formalizations of the replica 
method is still an open issue. 

\section*{Acknowledgments}
Numerous discussions with V. E. Kravtsov are appreciated with thanks.

\appendix
\widetext
\def\theequation{A.\arabic{equation}}
\section*{Equations (3.27) and (3.42)}
(i) Consider the average in the l.h.s. of Eq. (\ref{gauss-4a}), in which, by definition,
\def\theequation{A.\arabic{equation}}
\begin{eqnarray}
{\overline \Psi}^{\alpha}(x) &=& (-1)^{\alpha-1} \sum_{k=1}^{2N} \chi_k ^* 
\varphi_{k-1}^{(\alpha-1)}(x), \label{fe} \\
\Psi^{\beta}(y) &=& \sum_{k=1}^{2N} \chi_k \varphi_{k-1}^{(2-\beta)}(y) \label{se},
\end{eqnarray}
$[\alpha,\beta=1,2]$, see Eqs. (\ref{gf-4}) and (\ref{l-m}). To compute the required average given by
Eq. (\ref{aver-4}), we write
\begin{eqnarray}
\langle\langle {\overline \Psi}^{\alpha}(\sigma_1\Delta_N)
  \Psi^{\beta}(\sigma_2 \Delta_N) \rangle\rangle = (-1)^{\alpha-1}
\sum_{j,k=1}^{2N} \langle \chi_j^* \chi_k \rangle
\varphi_{j-1}^{(\alpha-1)}(\sigma_1\Delta_N) \varphi_{k-1}^{(2-\beta)}
(\sigma_2\Delta_N) , \label{a1-4}
\end{eqnarray}
where
\begin{eqnarray}
\label{eq-chi-1}
\langle \chi_j^* \chi_k \rangle = \frac{\int \left(\prod_{m=1}^{2N} 
d\chi_m^* d\chi_m\right) 
\chi_j^* \chi_k e^{\sum_{p,q=1}^{2N} \chi_p^* M_{pq} \chi_q }}
{
\int \left(\prod_{m=1}^{2N} d\chi_m^* d\chi_m\right) 
e^{\sum_{p,q=1}^{2N} \chi_p^* M_{pq} \chi_q }
},
\end{eqnarray}
and $M_{pq}$ is specified by Eq. (\ref{m-matrix-4}). Since the Grassmann integral
Eq. (\ref{eq-chi-1}) equals
$(M^{-1})_{kj} \equiv \mu_{kj}$, one derives from Eq. (\ref{a1-4}):
\begin{eqnarray}
\langle\langle {\overline \Psi}^{\alpha}(\sigma_1\Delta_N)
  \Psi^{\beta}(\sigma_2 \Delta_N) \rangle\rangle = (-1)^{\alpha-1}
\sum_{j,k=1}^{2N} 
\varphi_{j-1}^{(\alpha-1)}(\sigma_1\Delta_N) 
\mu_{kj} \varphi_{k-1}^{(2-\beta)}
(\sigma_2\Delta_N)
= {\hat K}_N^{\beta \alpha}
(\sigma_2\Delta_N, \sigma_1\Delta_N). \label{a1-41}
\end{eqnarray}
This completes the proof of Eq. (\ref{gauss-4a}).

(ii) Consider the average in the l.h.s. of Eq. (\ref{gauss-4b}). Making use of 
Eqs. (\ref{fe}) and (\ref{se}), we rewrite the average as
\begin{eqnarray}
\label{4f4}
\langle\langle 
{\overline \Psi}^{\alpha}(\sigma_1\Delta_N) \Psi^{\beta}(\sigma_2\Delta_N )
{\overline \Psi}^{\gamma}(\sigma_3\Delta_N) \Psi^{\delta}(\sigma_4\Delta_N )  
\rangle \rangle
&=& (-1)^{\alpha+\gamma} \sum_{r,p,q,s=1}^{2N} 
\langle \chi_r^* \chi_p \chi_q^* \chi_s \rangle \nonumber \\
&\times& \varphi_{r-1}^{(\alpha-1)}(\sigma_1\Delta_N)
\varphi_{p-1}^{(2-\beta)}(\sigma_2\Delta_N) 
\varphi_{q-1}^{(\gamma-1)}(\sigma_3\Delta_N)
\varphi_{s-1}^{(2-\delta)}(\sigma_4\Delta_N),
\end{eqnarray}
where
\begin{eqnarray}
\label{4h4}
\langle \chi_r^* \chi_p \chi_q^* \chi_s \rangle = 
\frac{\int \left(\prod_{m=1}^{2N} d\chi_m^* d\chi_m\right) 
\chi_r^* \chi_p \chi_q^* \chi_s e^{\sum_{p,q=1}^{2N} \chi_p^* M_{pq} \chi_q }}
{
\int \left(\prod_{m=1}^{2N} d\chi_m^* d\chi_m\right) 
e^{\sum_{p,q=1}^{2N} \chi_p^* M_{pq} \chi_q }
} \equiv \mu_{pr}\mu_{sq} - \mu_{sr}\mu_{pq}.
\end{eqnarray}
Inserting this result into Eq. (\ref{4f4}) and appealing to Eq. (\ref{k-44}), 
we arrive at the r.h.s. of Eq. (\ref{gauss-4b}). This completes the proof.

(iii) Consider the average in the l.h.s. of Eq. (\ref{w-1a}) in notations of Eq. (\ref{gauss-1}). 
By definition,
\begin{eqnarray}
{\overline \Psi}^\alpha(x) &=& (-1)^{\alpha-1}\sum_{k=1}^{2N} \chi_k^* 
[{\hat \epsilon}^{2-\alpha}\varphi_{k-1}](x), 
\label{def-1} \\
\Psi^\beta (y) &=& \sum_{k=1}^{2N} \chi_k [{\hat \epsilon}^{\beta-1}\varphi_{k-1}](y),
\label{def-2} 
\end{eqnarray}
$[\alpha,\beta=1,2]$, see Eq. (\ref{tcgf-1}). The required average is
\begin{eqnarray}
\label{av-1ap}
\langle\langle {\overline \Psi}^\alpha (\sigma_1\Delta_{2N})
\Psi^\beta (\sigma_2\Delta_{2N}) \rangle\rangle  = (-1)^{\alpha-1}\sum_{j,k=1}^{2N}
\langle \chi_j^* \chi_k\rangle [{\hat \epsilon}^{2-\alpha} \varphi_{j-1}](\sigma_1\Delta_{2N})
[{\hat\epsilon} ^{\beta-1}\varphi_{k-1}](\sigma_2\Delta_{2N}),
\end{eqnarray}
where $\langle \chi_j^* \chi_k \rangle$ is given by Eq. (\ref{eq-chi-1}) with
$M_{pq}$ defined by Eq. (\ref{M-1}). We then conclude that $\langle\chi_j^*\chi_k\rangle$
equals $(M^{-1})_{kj}=\mu_{kj}$, so that
\begin{equation}
\label{av-2-146}
\langle\langle
{\overline \Psi}^\alpha (\sigma_1\Delta_{2N}) \Psi^\beta (\sigma_2\Delta_{2N})
\rangle\rangle = (-1)^{\alpha-1} 
\sum_{j,k=1}^{2N}
[{\hat \epsilon}^{2-\alpha} \varphi_{j-1}](\sigma_1\Delta_{2N}) \mu_{kj}
[{\hat\epsilon} ^{\beta-1}\varphi_{k-1}](\sigma_2\Delta_{2N}) =
-{\hat Q}_{2N}^{\beta\alpha}(\sigma_2\Delta_{2N},\sigma_1\Delta_{2N}).
\end{equation}
Here, we have used Eq. (\ref{Q-eq}). This completes the proof of Eq. (\ref{w-1a}).

The proof of Eq. (\ref{w-1b}) is straightforward and goes along the lines of (ii)
and (iii).
\narrowtext

\widetext

\begin{thebibliography}{10}
\bibitem{Remark-unitary} This formula is used for $\beta=2$. For technical reasons, 
at $\beta=1$ and $\beta=4$ we make use of Eq. (\ref{rl4b=4}).
\bibitem{Remark-lr4b=2} The simplest way to do so is to notice that $\varphi_k (x)$
is simultaneously the left and the right eigenfunction of the operator $K_N(x,y)$, i. e. 
$\int dy K_N(x,y) \varphi_k(y) = \varphi_k(x)$ and 
$\int dx \varphi_k (x) K_N(x,y) = \varphi_k(y)$ for $k \in (0,\ldots,N-1)$.
\bibitem{Remark-Wick} The identities Eqs. (\ref{wick-2}) can be proven along the
lines of the Appendix where a bit more involved cases of symplectic and orthogonal
symmetries are treated.
\bibitem{Remark1} The confinement potentials $V(E)$ is said to be `strong' if it 
increases at least as $|E|$ at infinity.
\end{thebibliography}
\end{document}